\documentclass[prl,twocolumn,superscriptaddress,showpacs,floatfix,longbibliography]{revtex4-1}
\usepackage{mathrsfs,braket}
\usepackage{amssymb, amsbsy, amsmath, latexsym, dsfont, array, layout,
graphicx,mathrsfs,color,ulem,bm}
\usepackage[colorlinks=true,citecolor=blue,urlcolor=blue]{hyperref}

\newcommand{\one}{\mathds{1}}

\begin{document}

\title{Observation of non-Hermitian topological Anderson insulator in quantum dynamics}

\author{Quan Lin}\thanks{These authors contributed equally to this work}
  \affiliation{Beijing Computational Science Research Center, Beijing 100084, China}
\author{Tianyu Li}\thanks{These authors contributed equally to this work}
  \affiliation{CAS Key Laboratory of Quantum Information, University of Science and Technology of China, Hefei 230026, China}
  \affiliation{CAS Center For Excellence in Quantum Information and Quantum Physics, Hefei 230026, China}
\author{Lei Xiao}
  \affiliation{Beijing Computational Science Research Center, Beijing 100084, China}
\author{Kunkun Wang}
  \affiliation{Beijing Computational Science Research Center, Beijing 100084, China}
\author{Wei Yi}\email{wyiz@ustc.edu.cn}
  \affiliation{CAS Key Laboratory of Quantum Information, University of Science and Technology of China, Hefei 230026, China}
 \affiliation{CAS Center For Excellence in Quantum Information and Quantum Physics, Hefei 230026, China}
\author{Peng Xue}\email{gnep.eux@gmail.com}
  \affiliation{Beijing Computational Science Research Center, Beijing 100084, China}

\begin{abstract}
\bf{Disorder and non-Hermiticity dramatically impact the topological and localization properties of a quantum system, giving rise to
intriguing quantum states of matter. The rich interplay of disorder, non-Hermiticity, and topology is epitomized
by the recently proposed non-Hermitian topological Anderson insulator that hosts a plethora of exotic phenomena.
Here we experimentally simulate the non-Hermitian topological Anderson insulator using disordered photonic quantum walks,
and characterize its localization and topological properties. In particular, we focus on the competition between Anderson localization induced by random disorder, and the non-Hermitian skin effect under which all eigenstates are squeezed toward the boundary. The two distinct localization mechanisms prompt a non-monotonous change in profile of the Lyapunov exponent, which we experimentally reveal through dynamic observables. We then probe the disorder-induced topological phase transitions, and demonstrate their biorthogonal criticality. Our experiment further advances the frontier of synthetic topology in open systems.}
\end{abstract}

\maketitle

Topological edge states in topological materials are robust against weak perturbations,
an ability originating from the global geometry of eigen wave functions in the Hilbert space~\cite{kane,Qi}. Such an intrinsic geometric feature is captured by global topological invariants that are related to edge states through the bulk-boundary correspondence. However, this conventional paradigm is challenged by localization under disorder~\cite{shen,been,Franz,Hughes} or non-Hermiticity~\cite{Lee,WZ1,WZ2,Budich,mcdonald,alvarez,murakami,ThomalePRB,fangchenskin,kawabataskin,Slager,XZ,yzsgbz,stefano,tianshu,lli}, which have become the focus of study of late, particularly in light of recent experimental progresses in synthetic topological systems~\cite{sciencecd,teskin,teskin2d,metaskin,photonskin,scienceskin,EP,dzou,Szameit,Szameit1,Szameit2}. On one hand, despite its gap-closing tendency, disorder can induce topology from a trivial insulator. In the resulting topological Anderson insulator, the global topology is carried by fully localized states in the bulk~\cite{shen,been,Franz,Hughes}. On the other hand, in a broad class of non-Hermitian topological systems, the nominal bulk states are exponentially localized toward boundaries under the non-Hermitian skin effect~\cite{WZ1,WZ2,Budich,mcdonald,alvarez,murakami,ThomalePRB,fangchenskin,kawabataskin,Slager,XZ,yzsgbz,stefano,tianshu,lli}. The deviation of the bulk-state wave functions from the extended Bloch waves invalidates the conventional bulk-boundary correspondence, necessitating the introduction of non-Bloch topological invariants~\cite{WZ1,WZ2,murakami,ThomalePRB}. While the two localization mechanisms differ in origin and manifestation, topology of the underlying system gets fundamentally modified in either case. Remarkably, in the recently proposed non-Hermitian topological Anderson insulator~\cite{ZDW1,ZDW2,ZCW,hug}, the two distinct localization mechanisms are pitted against each other, wherein the interplay of disorder, non-Hermiticity, and topology leads to exotic phenomena such as the non-monotonous localization, disorder-induced non-Bloch topological phase transitions, and biorthogonal critical behaviors.

In this work, we report the experimental observation of non-Hermitian topological Anderson insulators in single-photon quantum-walk dynamics.
Driven by a non-unitary topological Floquet operator, the quantum walk undergoes polarization-dependent photon loss, and acquires the non-Hermitian skin effect.
In contrast to previously implemented quantum walks with the non-Hermitian skin effect~\cite{photonskin,EP}, our current experiment resorts to the time-multiplexed configuration, with the spatial degrees of freedom encoded in the discrete arrival time of photons at the detector~\cite{Silb1}. This enables us to implement quantum walks with a larger number of time steps, which is pivotal for the current experiment.
We introduce static random disorder through parameters of the optical elements~\cite{Silb2}, which would result in a complete localization of bulk states in the large-disorder limit. In the intermediate regime with moderate loss and disorder,
the competition between the non-Hermitian skin effect and Anderson localization yields non-monotonic localization features which we characterize by measuring the Lyapunov exponent~\cite{stefano}.
Using the biorthogonal chiral displacement, we then probe the topological phase transition, which is in qualitatively agreement with theoretical predictions. At the measured topological phase boundary, the biorthogonal localization length diverges, consistent with the biorthogonal critical nature of the phase transition~\cite{ZDW1,ZDW2,ZCW}.

\begin{figure*}
\centering
\includegraphics[width=0.75\textwidth]{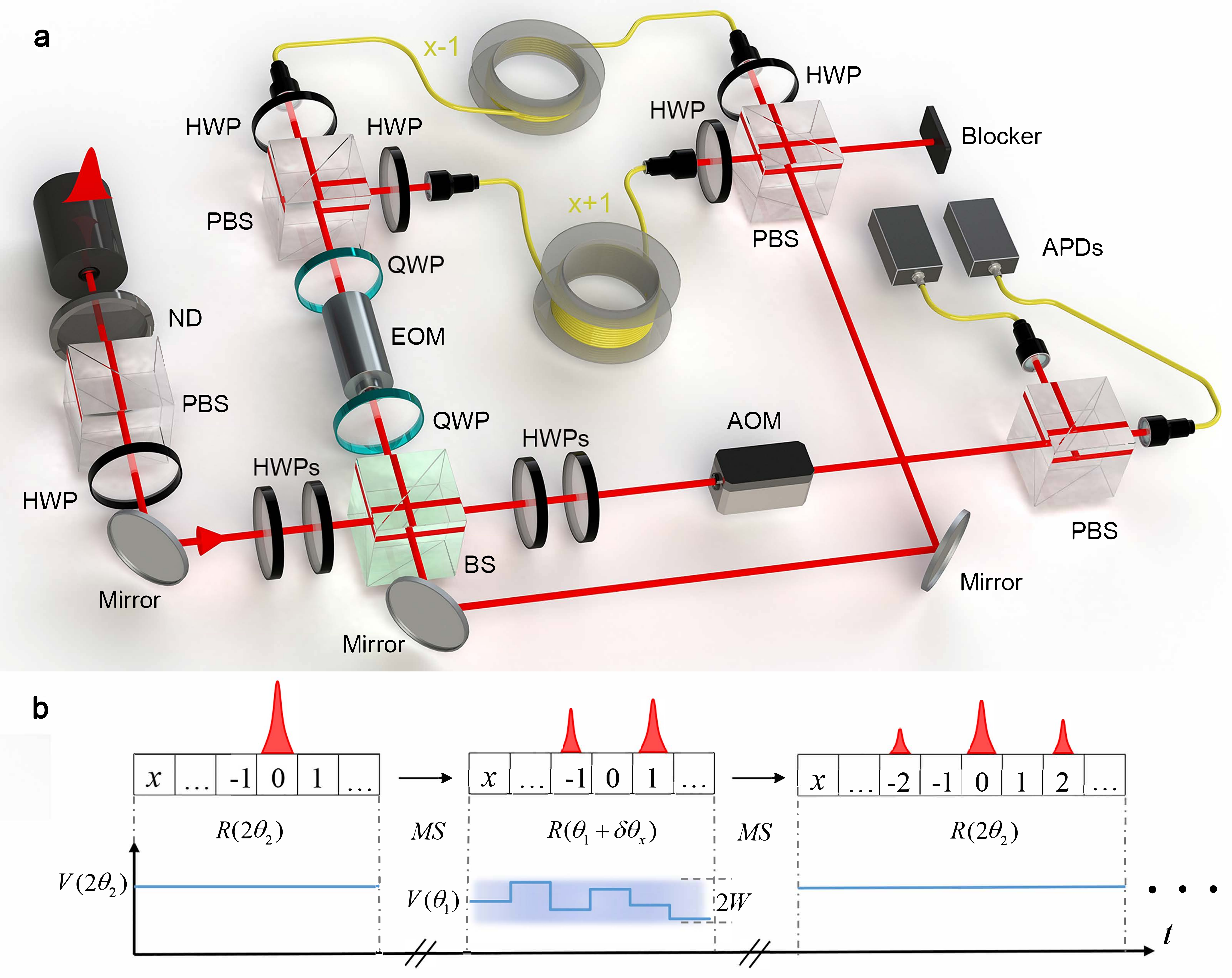}
\caption{{\bf Experimental setup for observing non-Hermitian topological Anderson insulator.} {\bf{a}}, Photons are coupled in and out of an interferometric network through a low-reflectivity beam splitter (BS, reflectivity $5\%$). The coin operation is carried out with wave plates and a dynamic electro-optic modulator (EOM). The shift operator is realized by splitting the light, using a polarizing beam splitter (PBS), into two single-mode fibers of length of $160.000$m and $167.034$m, respectively. As such, the spatial modes are encoded into the polarization-dependent temporal shift within a time step. The out-coupled photons are detected using the avalanche photo-diodes (APDs), in a time- and polarization-resolved fashion. An optical switch acousto-optic modulator (AOM) is used to protect the APDs such that photons are only allowed to reach the APDs at the time of measurement. {\bf{b}}, Illustration of the operation sequence of the time-multiplexed quantum walk. Here $V(\theta_i)$ is the control voltage applied to the EOM for generating rotations with the coin parameter $\theta_i$.
}
\label{fig:fig1}
\end{figure*}

{\bf Results}

{\bf A time-multiplexed non-unitary quantum walk.}
We implement a one-dimensional photonic quantum walk governed by the Floquet operator
\begin{align}
U=R(\theta_2)MSR(\theta_1)MSR(\theta_2).
\label{eq:U}
\end{align}
Here the shift operator is given by
$S=\sum_x \ket{x-1}\bra{x}\otimes\ket{H}\bra{H}+\ket{x+1}\bra{x}\otimes\ket{V}\bra{V}$, with $|H\rangle$ ($|V\rangle$) the horizontally (vertically) polarized state. The non-unitary operator $M=\sum_x\ket{x}\bra{x}\otimes
\begin{pmatrix}
e^{\gamma} & 0 \\
0 & e^{-\gamma}\end{pmatrix}$ with $\gamma$ the gain-loss parameter.
The coin operator $R(\theta)
=\sum_x\ket{x}\bra{x}\otimes\begin{pmatrix}
\cos\theta & -\sin\theta \\
\sin\theta & \cos\theta
\end{pmatrix}$, where the matrix is in the basis $\{|H\rangle,|V\rangle\}$. The quantum walk governed by $U$ features the non-Hermitian skin effect (see Supplemental Material), which we demonstrate later with dynamic measurements.

For the experimental implementation, we adopt a time-multiplexed scheme, as illustrated in Fig.~\ref{fig:fig1}. Photons are sent through an interferometric network consisting of optical elements for a half step of the discrete-time quantum walk in Eq.~(\ref{eq:U}).
The shift operator is implemented by separating the two polarization components and routing them through fibres of different lengths to introduce polarization-dependent time delay, such that the walker position is mapped to the time domain. For instance, a superposition of multiple spatial positions at a given time step is translated into the superposition of multiple well-resolved pulses within the same discrete time step.
A pair of wave plates are introduced into each of the paths, to realize a polarization-dependent loss operation $M_\text{E}=\sum_x\ket{x}\bra{x}\otimes \left(|H\rangle\langle H|+e^{-2\gamma}|V\rangle\langle V|\right)$, which is related to $M$ through $M=e^{\gamma} M_\text{E}$. We therefore read out the time-evolved state driven by $U$ by adding a time-dependent factor $e^{\gamma t}$ to our experimental measurement.
To implement the coin operator, an electro-optical modulator (EOM) is inserted into the main interferometric cycle, in combination with wave plates, to provide a carefully time-sequenced control over $\theta$. Importantly, the EOM enables an individual-pulse-resolved coin operation, providing the basis for the implementation of a walker-position-dependent disorder. The disorder is introduced to the operator $R(\theta_1)$ in Eq.~(\ref{eq:U}), where the actual rotation angle modulated by a small position-dependent $\delta\theta(x)$, with $\delta\theta(x)$ randomly taking values within the range of $\left[-W,W\right]$. Here $W$ indicates the disorder strength. We implement only static disorder for our experiments, such that $\delta\theta(x)$ does not change with time steps.

For the input and out-coupling of the interferometric network, a beam splitter (BS) with a reflectivity of $5\%$ is introduced, corresponding to a low coupling rate of photons into the network, but also enabling the out-coupling of photons for measurement. For that purpose, two avalanche photo-diodes (APDs) are employed to record the out-coupled photons' temporal and polarization properties, yielding information regarding the number of time steps, as well as the spatial and coin states of the walker.

\begin{figure}
\includegraphics[width=0.5\textwidth]{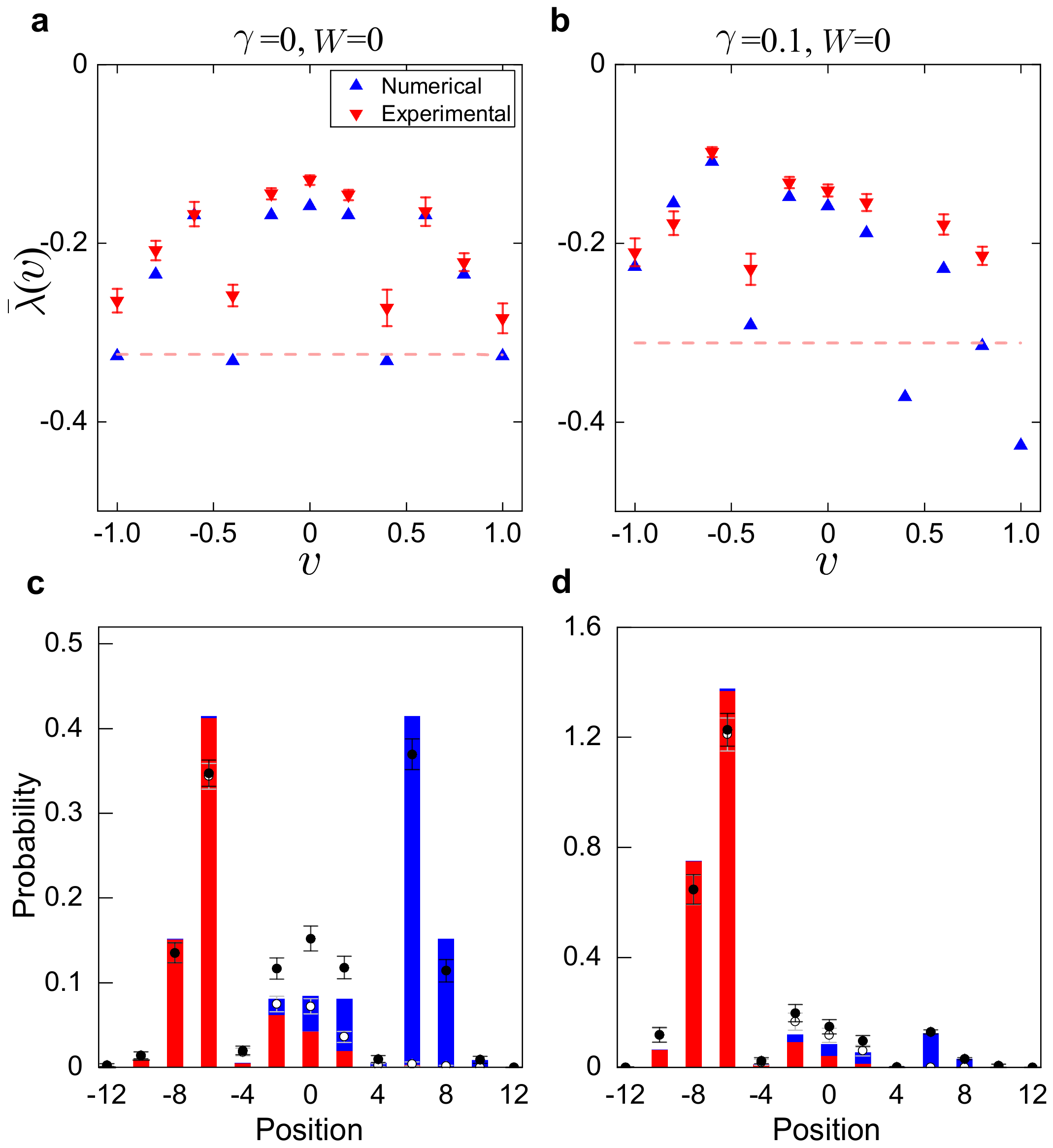}
\caption{{\bf Non-Hermitian skin effect from the Lyapunov exponent.} Measured polarization-averaged growth rates $\bar{\lambda}(v)$ for a unitary quantum walk with $\gamma=0$ in {\bf{a}}, and a non-unitary quantum walk with $\gamma=0.1$ in  {\bf{b}}. Red triangles with error bars are the experimental data, and blue triangles are from numerical simulations. The horizontal dashed line indicates the threshold values below which experimental data are no longer reliable due to photon loss. To construct $\bar{\lambda}$, we initialize the walker in the state $|0\rangle\otimes |H\rangle$ ($|0\rangle\otimes |V\rangle$), evolve it up to $10$ steps under the parameters $(\theta_1=4.3, \theta_2=2.175, W=0)$, and projectively measure the horizontally (vertically) polarized photon distribution following the last time step. We construct $\lambda_H$ and $\lambda_V$ from these polarization-resolved distributions, from which $\bar{\lambda}$ is calculated.
{\bf{c}} and {\bf{d}}: The polarization-resolved photon distribution after the last time step $t=10$, for the dynamics in {\bf{a}} and {\bf{b}}, respectively. Blue and red bars are respectively the numerical results for the horizontal-polarization-resolved and vertical-polarization-resolved photon distributions. White dots are the experimental measurements for the vertical-polarization-resolved photon distribution, and black dots are the experimental results for the sum of the polarization-resolved distributions.  Error bars are due to the statistical uncertainty in photon-number-counting.
}
\label{fig:fig2}
\end{figure}

{\bf Non-Hermitian skin effect.}
While a defining signal of the non-Hermitian skin effect is the accumulation of eigen wave functions at the boundary, it also impacts dynamics in the bulk, leaving unique signatures in the Lyapunov exponent.
Here the Lyapunov exponent is defined as
$\lambda(v)= \displaystyle{\lim_{t\rightarrow \infty}} \frac{1}{t}\text{log} \Big|\psi(x=vt,t)\Big|$~\cite{stefano},
where $v$ is the shift velocity, and $\psi(x,t)$ is the wave-function component at position $x$ and time step $t$.
Remarkably, for a system with the non-Hermitian skin effect, $\lambda(v)$ takes a maximum value at $v\neq 0$ for bulk dynamics far away from any boundary~\cite{stefano}.
By contrast, in the absence of the non-Hermitian skin effect,
$\lambda(v)$ acquires a symmetric profile with respect to its peak at $v=0$.
Such a behavior is closely related to the non-vanishing bulk current associated with the non-Hermitian skin effect, as well as the
accumulation of population at boundaries should they become relevant.

For our experiment, we implement $10$-step quantum walks without imposing any boundary or domain, and measure the polarization-averaged growth rate
\begin{align}
\bar{\lambda}(v)=\frac{\lambda_H(v)+\lambda_V(v)}{2}. 
\label{eq:avgL}
\end{align}
Here the additional average over polarization enables us to qualitatively capture the distinctive features of the Lyapunov exponent using a relatively small number of time steps ($t=10$).
In Eq.~(\ref{eq:avgL}), the polarization-resolved growth rates are defined as
$\lambda_i(v)=\frac{1}{t}\log|\psi^{(i)}_{x=vt}|$. To construct $\psi^{(i)}_{x=vt}=\langle i|\otimes\langle x| U^t |0\rangle \otimes|i\rangle$ ($i=H,V$), we initialize the walker in the state $|0\rangle\otimes|i\rangle$, and projectively measure the probability distribution of photons in the polarization state $|i\rangle$ of the spatial mode $|x\rangle$, following the last time step ($t=10$).
Note that the average over polarization in Eq.~(\ref{eq:avgL}) is taken for a faster convergence of the growth rate at a finite evolution time to the Lyapunov exponent.

In Fig.~\ref{fig:fig2}, we show the measured polarization-averaged growth rates as functions of the shift velocity, for {\bf a}, {\bf c} the unitary, and {\bf b}, {\bf d} the non-unitary cases, both without disorder. Apparently, under the non-Hermitian skin effect ($\gamma\neq 0$), the peak of the growth rate lies with a finite $v$ (Fig.~\ref{fig:fig2}{\bf{b}}), in contrast to the more symmetric profile without skin effect (Fig.~\ref{fig:fig2}{\bf{a}}). Such a growth-rate profile directly originates from the directional propagation of probability in the bulk, as clearly indicated in the measured polarization-resolved probability distributions after the final time step (Figs.~\ref{fig:fig2}{\bf c} and {\bf d}). In the presence of open boundaries, the directional probability propagation naturally leads to the accumulation of population at the boundaries.

\begin{figure*}
\includegraphics[width=0.75\textwidth]{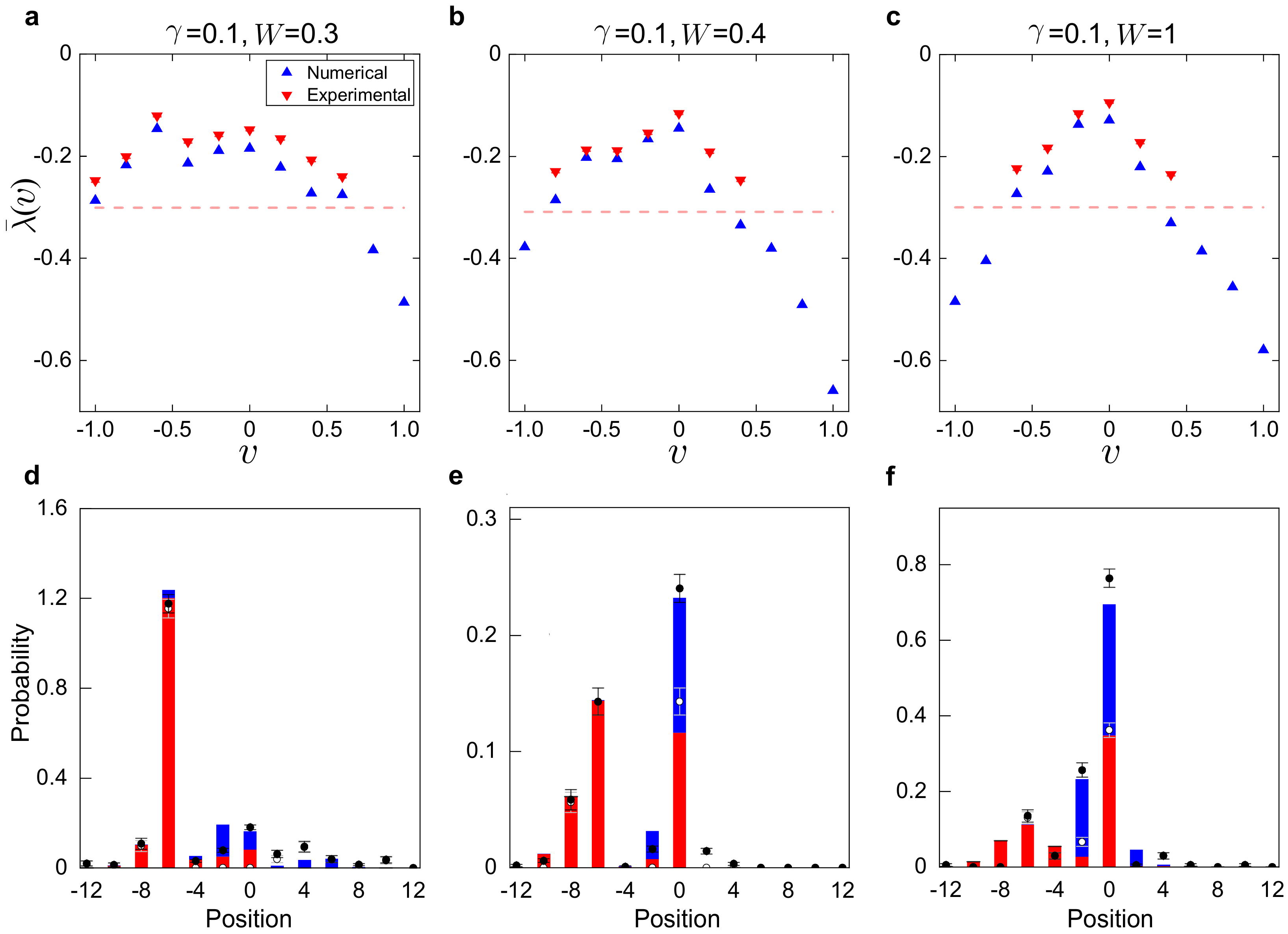}
\caption{{\bf Competition between the non-Hermitian skin effect and Anderson localization.} {\bf a}, {\bf b}, {\bf c}, The measured $\bar{\lambda}(v)$ with increasing $W$, under the parameters $\theta_1=4.3$, $\theta_2=2.175$, and $\gamma=0.1$. {\bf d}, {\bf e}, {\bf f}, The experimental data (symbols) and numerical results (bars) for the polarization-resolved photon distribution after the final step $t=10$. In {\bf a}, {\bf b} and {\bf c}, we average over $20$ disorder configurations. Symbols used are the same as those in Fig.~\ref{fig:fig2}. Error bars are due to the statistical uncertainty in photon-number-counting.
}
\label{fig:fig3}
\end{figure*}

{\bf Competition with Anderson localization.}
We now switch on disorder, and investigate the interplay between the non-Hermitian skin effect and disorder~\cite{ZDW1,ZCW}. In Fig.~\ref{fig:fig3}, we show the measured $\bar{\lambda}(v)$ for increasing disorder strength $W$, under a fixed non-Hermitian parameter $\gamma$. When $W$ is small, the asymmetric profile persists (see Figs.~\ref{fig:fig3}{\bf a} and {\bf d}), indicating the dominance of the non-Hermitian skin effect. A careful comparison between Fig.~\ref{fig:fig2}{\bf{b}} and Fig.~\ref{fig:fig3}{\bf{a}} suggests the emergence of another peak at $v=0$, though only just visible in Fig.~\ref{fig:fig3}{\bf{a}}. The peak at $v=0$ rapidly rises with increasing $W$.
This leads to the twin-peak structure under an intermediate $W$, as shown in Figs.~\ref{fig:fig3}{\bf b} and {\bf e}.
This is a direct evidence for the competition between the disorder induced Anderson localization and the non-Hermitian skin effect.
Finally, for sufficiently large $W$, $\bar{\lambda}(v)$ again peaks at $v=0$, as Anderson localization completely suppresses probability flow in the bulk that leads to the non-Hermitian skin effect. Such a competition as revealed by our experiment is consistent with the non-monotonous localization predicted in Ref.~\cite{ZDW1}, where the inverse participation ratio is used to characterize the competition (see Supplemental Material).

\begin{figure}[tbp]
\includegraphics[width=0.5\textwidth]{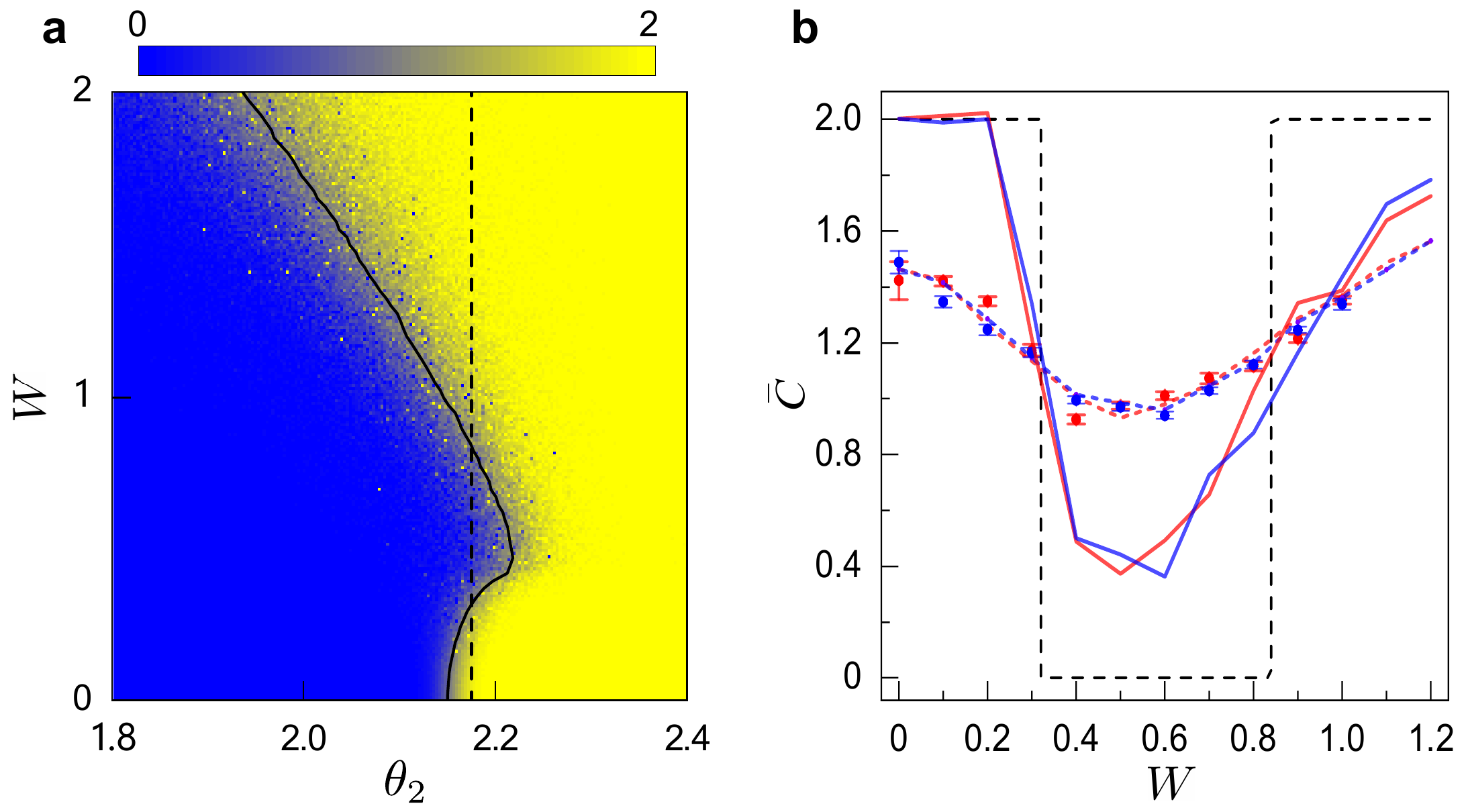}
\caption{{\bf Characterizing topology.} {\bf{a}}, Theoretical phase diagram in terms of the color contour of the numerically evaluated biorthogonal local marker, with $\theta_1=4.3+\delta \theta$ and $\gamma=0.1$. {\bf{b}}, Measured averaged chiral displacement for $9$-step quantum walks with $\theta_2=2.175$ (vertical dashed line in {\bf{a}}), averaged over $10$ different configurations of $\delta \theta(x)$, with $\delta\theta(x)$ taking random values within the range $\left[-W,W\right]$. Experimental data are represented by blue and red dots for $\gamma=0$ and $\gamma=0.1$, respectively.  Error bars are due to the statistical uncertainty in photon-number-counting. Blue and red dashed lines are numerically evaluated chiral displacements for $9$-step quantum walks, averaged over $2000$ random-disorder configurations, for $\gamma=0$ and $\gamma=0.1$, respectively. Blue and red solid lines are numerically evaluated chiral displacement for $400$-step quantum walks, averaged over $200$ random-disorder configurations, for $\gamma=0$ and $\gamma=0.1$, respectively. For all cases, the initial state is $|0\rangle\otimes|V\rangle$.}
\label{fig:fig4}
\end{figure}

{\bf Disorder-induced topology.}
The Floquet operator $U$ is topological, protected by the chiral symmetry with $\Gamma U\Gamma=U^{-1}$, where $\Gamma=\sum_x\ket{x}\bra{x}\otimes\sigma_x$. While the topology of $U$ generally persists under small random disorder, disorder can also induce non-trivial topology from a topologically trivial state, similar to the case with the topological Anderson insulator in Hermitian systems~\cite{shen,been,Franz,Hughes}.
In Fig.~\ref{fig:fig4}{\bf{a}}, we plot the theoretical phase diagram, characterized through the disorder-averaged local marker under the non-Bloch band theory (see Supplemental Material).
Incidentally, for our choice of $U$, the topological phase boundary is insensitive to $\gamma$, despite the presence of the non-Hermitian skin effect and the application of the non-Bloch band theory. Nevertheless, the biorthogonal localization length, rather than the conventional localization length, diverges at the topological phase boundary (solid black curve in Fig.~\ref{fig:fig4}{\bf{a}})~\cite{ZCW}, suggesting a unique non-Hermitian criticality.

Here we focus on the impact of disorder on the topological phase boundary, which
we experimentally probe through
the time and disorder averaged biorthogonal chiral displacement, defined for a $t$-step quantum walk as~\cite{ZCW,sciencecd}
\begin{align}
\bar{C}=\frac{1}{N}\sum_{n=1}^{N}\sum_{t'=1}^{t}\frac{1}{t}\langle \chi_n(t')|\Gamma X |\psi_n(t')\rangle,
\label{eq:C}
\end{align}
where $|\psi_n(t)\rangle =U^t|\psi(0)\rangle$ and $|\chi_n(t)\rangle=\left[(U^{-1})^\dag\right]^t|\psi(0)\rangle$, $|\psi(0)\rangle=|0\rangle\otimes|V\rangle$, the subscript $n$ indicates the $n$th disorder configuration (with a total of $N$ configurations), and $X$ is the position operator.
Experimentally, we prepare $|\psi_n(t)\rangle$ and $|\chi_n(t)\rangle$ by separately evolving the initial state with $U$ and $(U^{-1})^\dag$, followed by state tomography to reconstruct $|\psi_n(t)\rangle$ and $|\chi_n(t)\rangle$, respectively, before calculating $\bar{C}$ according to Eq.~(\ref{eq:C}).

In Fig.~\ref{fig:fig4}{\bf{b}}, we plot the measured $\bar{C}$. Similar to Ref.~\cite{sciencecd}, while the measured chiral displacement varies smoothly across the topological phase boundaries due to the limited number of time steps amenable to our experiment, it does show a tendency consistent with the theoretically predicted phase boundaries. Numerically, it is found that $\bar{C}$ approaches the topological invariants given by the local marker (dashed line) at much larger time steps.
Finally, the measured $\bar{C}$ is insensitive to $\gamma$, consistent with theoretical predictions using the local marker.

{\bf Discussion.}
We report the first experimental observation of a non-Hermitian topological Anderson insulator. Using dynamic observables, we demonstrate the two competing localization mechanisms inherent in the system, and reveal a disorder-induced topological phase transition.
Our experiment raises interesting theoretical questions as to the fate of localized states in a non-Hermitian many-body system with skin effect, as well as the interplay of non-Hermiticity, disorder and many-body interactions therein. On the application side, disorder and non-Hermiticity provide convenient control over key properties of non-Hermitian Anderson insulators, opening routes toward the design of tunable optical device for enhanced sensing and engineered quantum transport.

For future studies, it is hopeful to further increase the evolution time of the quantum-walk dynamics based on the time-multiplexed configuration, such that a more accurate determination of the Lyapunov exponent can be achieved. It would also be interesting to explore similar competitions for higher-dimensional non-Hermitian topological Anderson insulators.

{\bf Methods}

{\bf Experimental setup.}
To implement quantum walks governed by the Floquet operator $U$ in Eq.~(\ref{eq:U}), we adopt a time-multiplexed configuration, encoding the internal coin-state degrees of freedom in the photonic polarizations, and the external spatial modes in the discretized temporal shift  within a time step~\cite{Silb1}. The overall experimental configuration is illustrated in Fig.~\ref{fig:fig1}.

The wave packets of photons are generated by a pulsed laser source with a central wavelength of $808$nm, a pulse width of $88$ps, and a repetition rate of $31.25$kHz. The pulses are attenuated to the single-photon level using neutral density filters at the detection stage. For a unitary quantum walk, the probability that a photon undergoes a full round-trip without getting lost or detected is about $0.59$ per step and the detection efficiency is $0.03$ per step (taking into account the efficiency of APDs and the reflectivity of BSs). We ensure the average photon number per pulse at the detection stage to be less than $2\times 10^{-4}$, so that there is a negligible probability of a multi-photon event.

To implement $U$ with a fibre loop configuration, we rewrite the $t$-step time-evolution operator as $U^t=e^{2\gamma t}U^t_\text{E}$, where
\begin{align}
U^t_\text{E}&=\Big[R(\theta_2) M_{\text{E}}SR(\theta_1) M_{\text{E}}SR(\theta_2)\Big]^t\nonumber\\
&=R(\theta_2)U^t_\text{loop}R(-\theta_2),\label{eq:loop}
\end{align}
and
\begin{align}
U_\text{loop}=M_{\text{E}}SR(\theta_1) M_{\text{E}}SR(2\theta_2).\label{eq:Uloop}
\end{align}
Here the coin operator $R(\theta_i)$ and the shift operator $S$ are the same as those in Eq.~(\ref{eq:U}) and the ensuing discussions. The polarization-dependent loss operator $M_\text{E}=\sum_x\ket{x}\bra{x}\otimes \left(|H\rangle\langle H|+e^{-2\gamma}|V\rangle\langle V|\right)$, which is related to $M$ through $M=e^{\gamma} M_\text{E}$. For each cycle in the interferometric network, the walker state is subject to the operation $M_{\text{E}}SR(\theta)$, where $\theta$ is alternatingly modulated to be $2\theta_2$ or $\theta_1$ for odd or even cycles. As such, one cycle in the network roughly corresponds to a half step of the quantum walk. The coin operators $R(\theta_2)$ and $R(-\theta_2)$ are implemented at the input and out-coupling stage, respectively.

More specifically, the operator $R(-\theta_2)$ [$R(\theta_2)$] in Eq.~(\ref{eq:U}) is
implemented using two half-wave plates (HWPs) with setting angles $-\theta_2/2$ ($\theta_2/2$) and $0$, respectively, before (after) the photon is sent into (coupled out of) the network. For the input, photons are reflected by a low-reflectivity BS with a reflectivity $5\%$, such that there is a $5\%$ probability to couple a photon into the network. The same BS is subsequently used as the out-coupler, where photons, after completing cycles in the interferometer, have a $5\%$ probability of being reflected out of the cycle and into the detection module.

Within each interferometer cycle, the photon is first sent through a sandwich-type, QWP($0$)-EOM($4\theta_2$)-QWP($90^\circ$) configuration~\cite{lu}, which is used to implement the coin operator $R(2\theta_2)$ or $R(\theta_1)$ in Eq.~(\ref{eq:Uloop}). Here QWP is the abbreviation for quarter-wave plates.
The birefringent crystal inside the EOM is set at $45^\circ$ to the $x/y$ axis, so that the EOM acts on the photon polarizations as
$\tilde{R}_\text{EOM}(\vartheta)=\begin{pmatrix}
1 & 1 \\
-1 & 1
\end{pmatrix}
\begin{pmatrix}
e^{i\frac{\vartheta}{2}} & 0 \\
0 & e^{-i\frac{\vartheta}{2}}
\end{pmatrix}
\begin{pmatrix}
1 & -1 \\
1 & 1
\end{pmatrix}=\begin{pmatrix}
\cos\frac{\vartheta}{2} & i\sin\frac{\vartheta}{2} \\
i\sin\frac{\vartheta}{2} & \cos\frac{\vartheta}{2}
\end{pmatrix}$. The properties impose that $\phi_V(x)/\phi_H(x)=-1$. Thus, in combination with a pair of wave plates, an EOM can be used to modify the polarization of each pulse individually, providing the basis for realizing position-dependent coin operations $R(\vartheta)=\begin{pmatrix}
                                                          1 & 0 \\
                                                          0 & -i
                                                        \end{pmatrix}\tilde{R}_\text{EOM}(2\vartheta)\begin{pmatrix}
                                                          1 & 0 \\
                                                          0 & i
                                                        \end{pmatrix}=\begin{pmatrix}
                                                \cos\vartheta & -\sin\vartheta \\
                                                \sin\vartheta & \cos\vartheta
                                              \end{pmatrix}$.
For a disorder-free quantum walk, we sequence the EOM such that $\vartheta=2\theta_2$ for odd cycles and $\vartheta=\theta_1$ for even cycles.

The shift operator $S$ is implemented by separating different polarization components of a photon using polarizing beam splitters (PBSs) and routing them through fibers of different lengths to introduce a well-defined time delay in between. Specifically, horizontally polarized photons traverse the fiber loop in $751.680$ns, while vertical ones take $33.046$ns longer to complete the trip. The resulting temporal difference corresponds to a step in the spatial domain of $x\pm1$.
As such, each position in each time step is represented by a unique discrete-time bin, i.e., the position information is mapped into the time domain.

To implement the loss operator $M_{\text{E}}$, a pair of HWPs are inserted into each fibre loop, one at the entrance and one near the exit. Since the operator $M_{\text{E}}$ induces loss in the polarization state $\ket{V}$ with a probability $1-e^{-4\gamma}$, we adjust the setting angles of the HWPs, such that only the desired components are reflected (transmitted) by the PBS at the exit of the short (long) fibre loop into the blocker, rendering the dynamic within the main cycle non-unitary. 
We therefore read out the evolved states from our experiment with $M_\text{E}$ by adding a factor $e^{\gamma t}$.

At the output of the shift operator, the two paths are coherently recombined, and photons are sent back to the input BS for the next split-step. In order to realize a full time step, two cycles in the interferometer network are required, with the setting angle of the EOM alternating between $2\theta_2$ (odd cycle) and $\theta_1$ (even cycle).
We introduce static disorder to the coin operator $R(\theta_1)$ for odd cycles. This is achieved by modulating the setting of EOM by a small random amount $\delta\theta\in\left[-W, W\right]$ around $\theta_1$.
Here $\delta\theta$ is position dependent but time independent. Such static disorder preserves the chiral symmetry of $U$.

Finally, after a photon has completed multiple cycles and is coupled out of the network by the BS (with a probability of $5\%$), the coin operator $R(\theta_2)$ is applied, and the photon registers a click at an APD with a time jitter $350$ps for detection.

{\bf State tomography.}
For the detection of the time-averaged chiral displacement, we reconstruct the final state $|\psi(t)\rangle=U^t\ket{\psi(0)}$ and its left vector $\ket{\chi(t)}=\left[(U^{-1})^\dagger\right]^t\ket{\psi(0)}$ for each time step. Here we take the reconstruction of $\ket{\psi(t)}$ as an example. Since $U$ and the initial state $\ket{\psi(0)}=\ket{0}\otimes\ket{V}$ are purely real in the polarization basis $\{|H\rangle,|V\rangle\}$, we have the expansion
\begin{align}
\ket{\psi(t)}=\sum_x \Big[p_H(t,x)\ket{x}\otimes\ket{H}+p_V(t,x)\ket{x}\otimes\ket{V}\Big],
\label{eq:phiexp}
\end{align}
where the coefficients $p_\mu(t,x)$ ($\mu=H,V$) are also real.
Based on these, we perform three distinct measurements  $M_i$ ($i=1,2,3$) to reconstruct $|\psi(t)\rangle$ in the basis $\{|H\rangle,|V\rangle\}$. This amounts to measuring the absolute values and the r signs of the real coefficients $p_\mu(t,x)$, as we detail in the following.

First, we measure the absolute values $\left|p_\mu(t,x)\right|$. After the $t$th time step, photons in the position $x$ are sent to a detection unit $M_1$, which consists of a PBS and APDs. $M_1$ applies a projective measurement of the observable $\sigma_z$ on the polarization of photons. The counts of the horizontally polarized photons $N_H(t,x)$ and vertically polarized ones $N_V(t,x)$ are registered by the coincidences between one of the APDs in the detection unit, and the APD for the trigger photon. The measured probability distributions are
\begin{equation}
P_\mu(t,x)=\frac{e^{2\gamma}c(t)N_\mu(t,x)}{\sum_{x}\left[N_H(t,x)+N_V(t,x)\right]},
\end{equation}
where $c(t)=\text{Tr}\left[U^t_\text{E}|\psi(0)\rangle\langle\psi(0)|(U^\dagger_\text{E})^t\right]$.  The square root of the probability distribution $P_\mu(t,x)$ corresponds to $\left|p_\mu(t,x)\right|$.

Second, we determine the relative sign between the amplitudes $p_H(t,x)$ and $p_V(t,x)$ via the detection unit $M_2$, which consists of an HWP at $22.5^\circ$, a PBS and APDs. The only difference between $M_2$ and $M_1$ is the HWP at $22.5^\circ$, i.e., a projective measurement of the observable $\sigma_x$ on the polarization components of photons. The difference between the probability distributions of the horizontally and vertically polarized photons is given by
\begin{align}
P_H(t,x)-P_V(t,x)=2p_H(t,x)p_V(t,x),
\end{align}
which determines the relative sign between $p_H(t,x)$ and $p_V(t,x)$.

Third, we probe the relative sign between the amplitudes $p_H(t,x)$ and $p_V(t,x')$, which is necessary to calculate the summation of wave functions in different positions at each time step. We take the relative sign between the amplitudes in the positions $x$ and $x-2$ as an example. To this end, a detection unit $M_3$ is introduced, consisting of an extra loop, an HWP at $22.5^\circ$, a PBS and APDs. In the extra loop, the EOM is set to realize a rotation $R(\theta_2+3\pi/4)$. The horizontally polarized photons at both $x$ and $x-2$ are combined at the end of the loop. The projective measurement of the observable $\sigma_x$ is applied on the polarization components of photons via an HWP at $22.5^\circ$, a PBS and APDs. The difference between the probability distributions of the horizontally and vertically polarized photons is given by
\begin{align}
P_H(t,x)-P_V(t,x)=&-\left[p_H(t,x)+p_V(t,x)\right]\\
&\times\left[p_H(t,x-2)-p_V(t,x-2)\right].\nonumber
\end{align}
As we have determined the relative sign between $p_H(t,x)$ and $p_V(t,x)$ [between $p_H(t,x-2)$ and $p_V(t,x-2)$] with $M_2$, we determine, using $M3$, the relative sign between $p_\mu(t,x)$ and $p_\mu(t,x-2)$ for arbitrary $x$.

Note that, as the purpose of reconstructing the final state is to calculate the expectation value of the averaged chiral displacement, the global sign of $p_\mu(x,t)$ is unimportant.

{\bf Data availability}

Experimental data, any related experimental background information not mentioned in the text and other findings of this study are available from the corresponding author upon reasonable request.

{\bf Acknowledgments}

This work has been supported by the National Natural Science Foundation of China (Grant Nos. 12025401, U1930402 and 11974331). W. Y. acknowledges support from the National Key Research and Development Program of China (Grant Nos. 2016YFA0301700 and 2017YFA0304100). L. X. acknowledges support from the Project Funded by China Postdoctoral Science Foundation (Grant Nos. 2020M680006 and 2021T140045).

{\bf Author contributions}

Q. L. performed the experiments
with contributions from K. W. and L. X. W. Y. developed the theoretical aspects and performed the theoretical analysis with contribution from T. L., and wrote part of the paper. P. X. supervised the project, designed the experiments, analyzed the results and wrote part of the paper.

{\bf Competing interest declaration}

The authors declare no competing interests.

{\bf Additional information}

Correspondence and requests for materials should be addressed to Wei Yi (wyiz@ustc.edu.cn) and Peng Xue (gnep.eux@gmail.com)

\clearpage

\pagebreak
\widetext
\begin{center}
\textbf{\large  Supplemental Material for  ``Observation of non-Hermitian topological Anderson insulator in quantum dynamics''}
\end{center}
\setcounter{equation}{0}
\setcounter{figure}{0}
\setcounter{table}{0}
\makeatletter
\renewcommand{\theequation}{S\arabic{equation}}
\renewcommand{\thefigure}{S\arabic{figure}}
\renewcommand{\bibnumfmt}[1]{[S#1]}

In this Supplemental Material, we provide details on the theoretical characterization of $U$, as well as additional experimental data.

\section{The generalized Brillouin zone}

We briefly outline the calculation of the generalized Brillouin zone (GBZ), which is useful for later discussions. We first rewrite the Floquet operator
\begin{align}
U=\sum_{x}|x\rangle \langle x+2| \otimes A_1+ |x+2\rangle \langle x| \otimes A_2+ |x\rangle \langle x| \otimes A_3,
\end{align}
where
\begin{align}
A_1=&R(\theta_2)P_0MR(\theta_1)MP_0R(\theta_2),\\
A_2=&R(\theta_2)P_1MR(\theta_1)MP_1R(\theta_2),\\
A_3=&R(\theta_2)P_0MR(\theta_1)MP_1R(\theta_2)+R(\theta_2)P_1MR(\theta_1)MP_0R(\theta_2).
\end{align}
Here $P_0=|0\rangle\langle 0|$ and $P_1=|1\rangle\langle 1|$ are projectors. Note that for convenience, we denote the polarization state $|H\rangle$ ($|V\rangle$) as $|0\rangle$ ($|1\rangle$) throughout the Supplemental Material.

Following Refs.~\cite{photonskin,EP}, we write the bulk-state ansatz as $|\psi\rangle=\sum_x \beta^x|x\rangle\otimes|\phi\rangle$. From the eigen equation $U|\psi\rangle=\lambda|\psi\rangle$ ($\lambda$ is the eigenvalue), we have
\begin{align}
\beta^4+C(\lambda,\gamma,\theta_1,\theta_2)\beta^2+e^{-4\gamma}=0, \label{eq:GBZ}
\end{align}
where $C(\lambda,\gamma,\theta_1,\theta_2)$ is a c-number. Sorting the solutions as $|\beta_1|\leq |\beta_2|\leq|\beta_3|\leq|\beta_4|$, and requiring $|\beta_2|=|\beta_3|$, we get
\begin{align}
|\beta|=e^{-\gamma}.
\end{align}
Hence, the GBZ in this case is always circular on the complex plane, with its radius dependent on $\gamma$.

\section{Non-Hermitian skin effects of the quantum walk}

\begin{figure}
\centering
\includegraphics[width=0.8\textwidth]{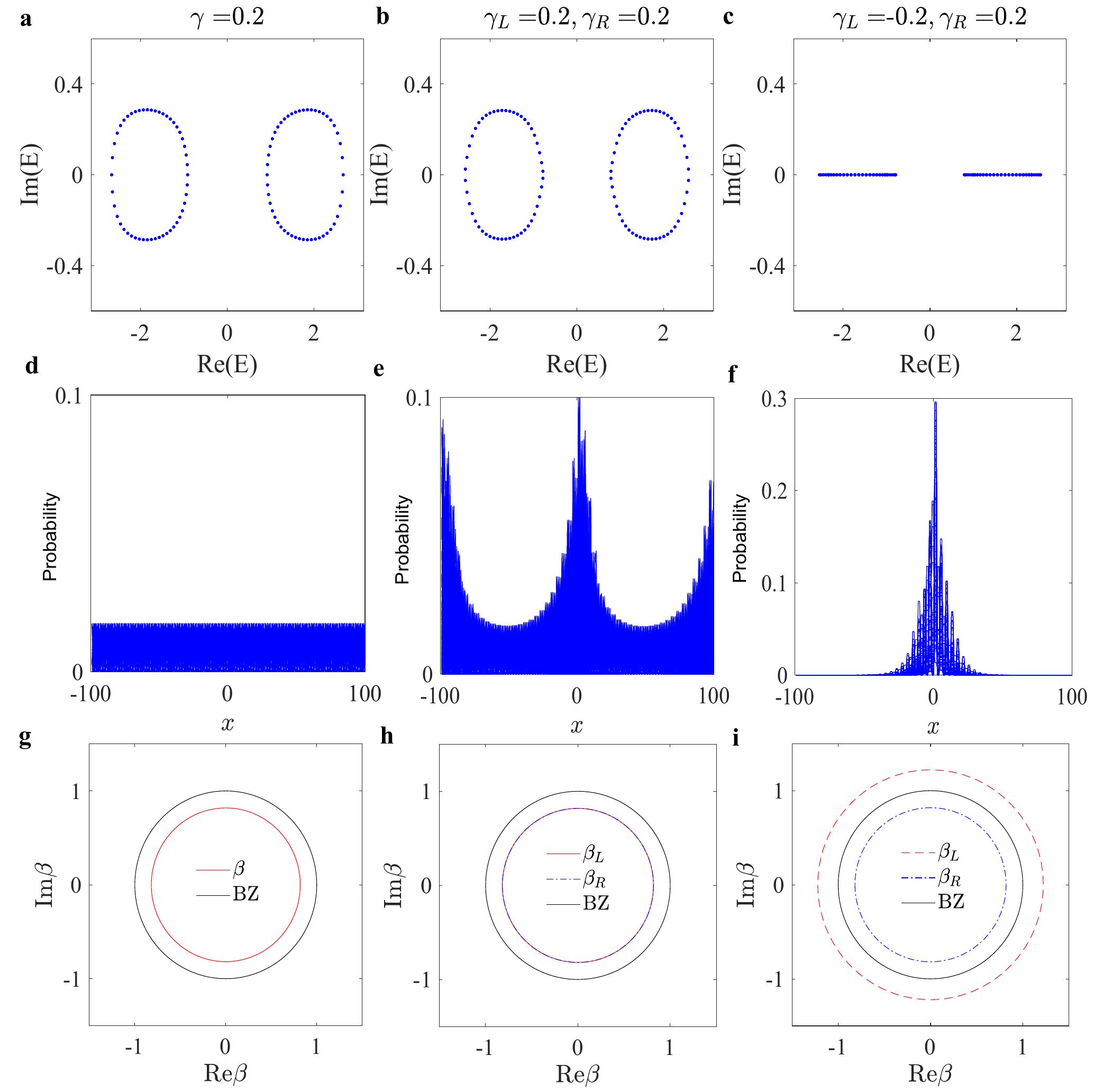}
\caption{Demonstration of the non-Hermitian skin effect. We take a domain-wall configuration with different parameters for $x<0$ and $x>0$. Therefore, two boundaries exist at $x=0$ and $x=100$. We show the quasienergy spectra (top panel), the spatial distribution of all eigen wave functions (middle panel), and the GBZs on the complex plane (lower panel).
{\bf a}, {\bf d}, {\bf g}, Periodic boundary condition with $\gamma=0.2$, $\theta_1=\pi/4$ and $\theta_2=0.1$. {\bf b}, {\bf e}, {\bf h}, Domain-wall configuration with
$\theta_{1}^L=\pi/4$, $\theta_{2}^L=0.1$, $\theta_{1}^R=\pi/4$, $\theta_{2}^R=\pi/2$, and $\gamma_L=\gamma_R=0.2$. The subscripts here indicate the left (L) and right (R) regions, respectively.
{\bf c}, {\bf f}, {\bf i}, Domain-wall configuration with the same coin parameters as those in {\bf b}, {\bf e} and {\bf h}, but different loss parameters $\gamma_L=-0.2$, $\gamma_R=0.2$.
For all calculations, we take the system size of $N_c=200$.
}
\label{fig:Sskin}
\end{figure}

The Floquet operator $U$ features non-Hermitian skin effect in the presence of boundaries. This is explicitly shown in Fig.~\ref{fig:Sskin}, where we calculate the quasienergy spectra (upper panel) and wave function distributions (lower panel) for the periodic boundary condition in {\bf a}, {\bf d}, {\bf g}; a domain-wall configuration with $|\beta_L|=|\beta_R|<1$ in {\bf b}, {\bf e}, {\bf h}; and a domain-wall configuration with $|\beta_R|<1<|\beta_L|$ in {\bf c}, {\bf f}, {\bf i}. For the domain wall configuration, the walker evolves along a ring, with $x\in\left[-100,100\right]$ and two boundaries located at $x=0$ and $x=100$. The left ($x<0$) and right ($x>0$) regions of the ring are characterized by different parameters in {\bf b}, {\bf e}, {\bf h} and {\bf c}, {\bf f}, {\bf i}, respectively.

Apparently, under the periodic boundary condition, the eigen wave functions are not localized. Whereas under the domain-wall configuration, the wave functions can be localized near one of the boundaries, or both. Intriguingly, despite the non-Hermitian skin effects, the eigenspectra in Fig.~\ref{fig:Sskin}{\bf b} still form closed loops on the complex plane, with non-trivial spectral winding. This is in contrast to Fig.~\ref{fig:Sskin}{\bf c}, where no spectral winding is present. The abnormal behavior of Fig.~\ref{fig:Sskin}{\bf b} calls for further study in the future.

\begin{figure}
\centering
\includegraphics[width=0.8\textwidth]{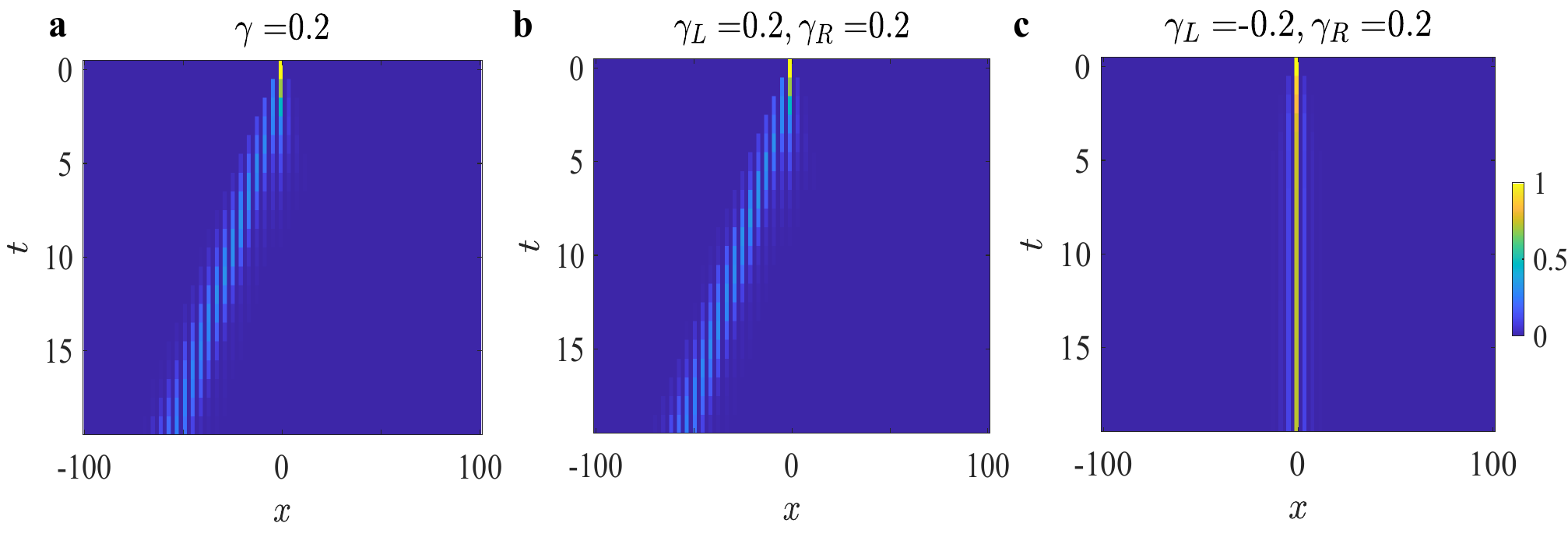}
\caption{Numerically simulated polarization-averaged population distribution for $20$-step quantum walks near the boundary at $x=0$ under the domain-wall configuration. The parameters for {\bf a}, {\bf b} and {\bf c} respectively correspond to those in Figs.~\ref{fig:Sskin}{\bf a}, {\bf b} and {\bf c}. For the dynamics, we initialize the walker in states
$|-1\rangle\otimes |0\rangle$ and $|-1\rangle\otimes |1\rangle$, respectively, independently evolve the two initial states under the same parameters, and take the average of the spatial probability distribution at each time step.
}
\label{fig:Spop}
\end{figure}

Further, in the case of Fig.~\ref{fig:Sskin}{\bf c}, the bulk population flows of the left and right regions are in the same direction, such that a bulk probability current persists even in the presence of boundaries. It follows that there is no dynamic population accumulation at the boundary, despite the presence of non-Hermitian skin effect. Nevertheless, dynamic population accumulation at the boundary can be observed by taking the parameters of Fig.~\ref{fig:Sskin}{\bf c}, where the bulk current no longer exists.
In Fig.~\ref{fig:Spop}, we show the numerically calculated polarization-averaged probability evolution for $20$-step quantum walks for these cases.

\section{Growth rates for longer time evolutions}

\begin{figure}
\centering
\includegraphics[width=0.8\textwidth]{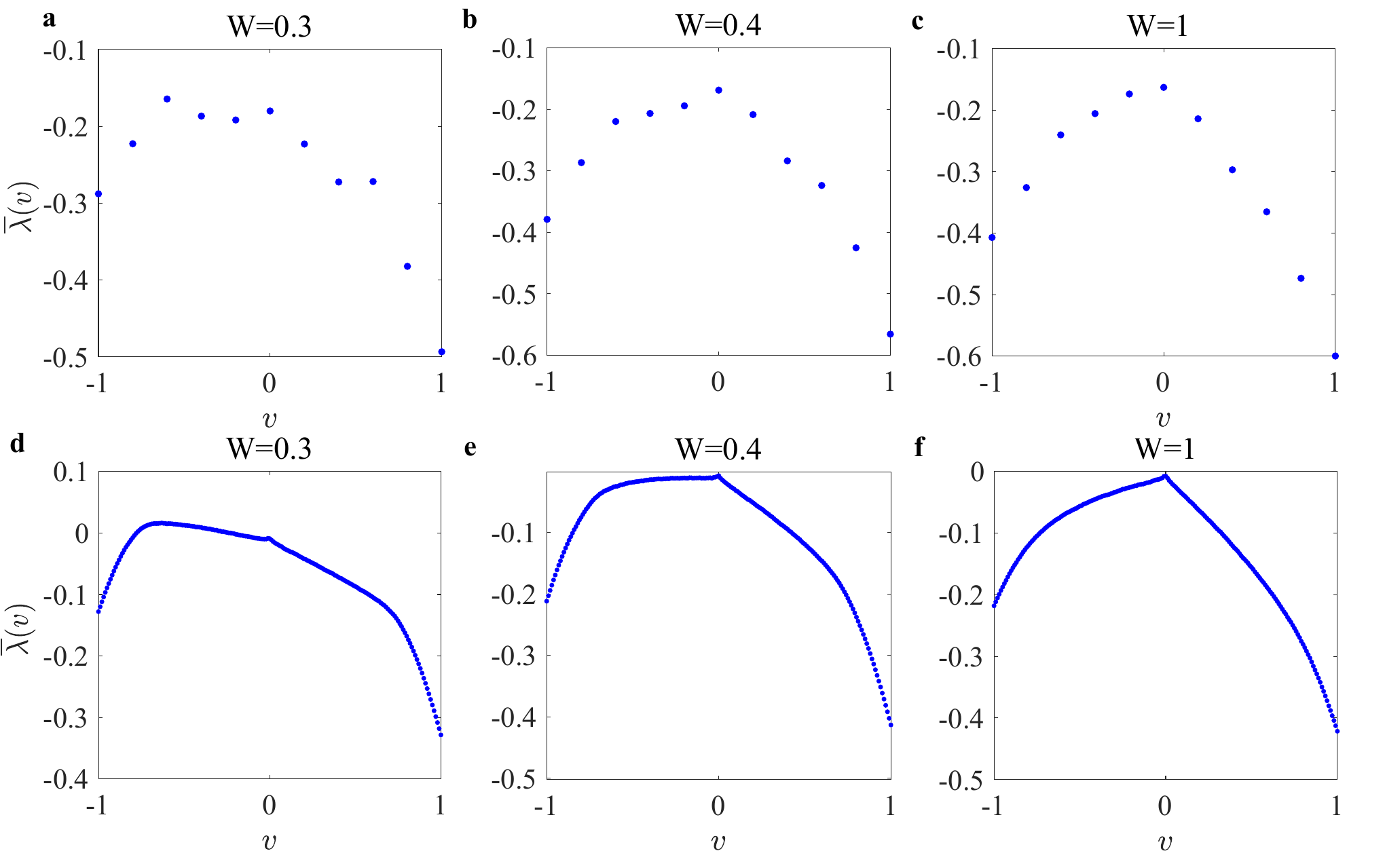}
\caption{Numerically calculated growth rates for quantum walks up to $10$ steps in {\bf a}, {\bf b} and {\bf c} and to $200$ steps in {\bf d}, {\bf e} and {\bf f}, respectively.
We take $\theta_{1}=4.3$, $\theta_{2}=2.175$, and $\gamma=0.1$. All data are averaged over $200$ disorder configurations.}
\label{fig:Slam}
\end{figure}

In Fig.~\ref{fig:Slam}, we show a comparison between the numerically calculated polarization-averaged growth rates for $10$-step quantum walks (upper panel), and $200$-step quantum walks (lower panel), respectively. The non-monotonic behavior of the growth rates with increasing $W$ is qualitatively the same for the short and long time dynamics. This enables us to experimentally probe the competition between the two localization mechanisms using experimentally achievable time steps.

\section{Characterizing localization through the inverse participation ratio}

\begin{figure}
\centering
\includegraphics[width=0.4\textwidth]{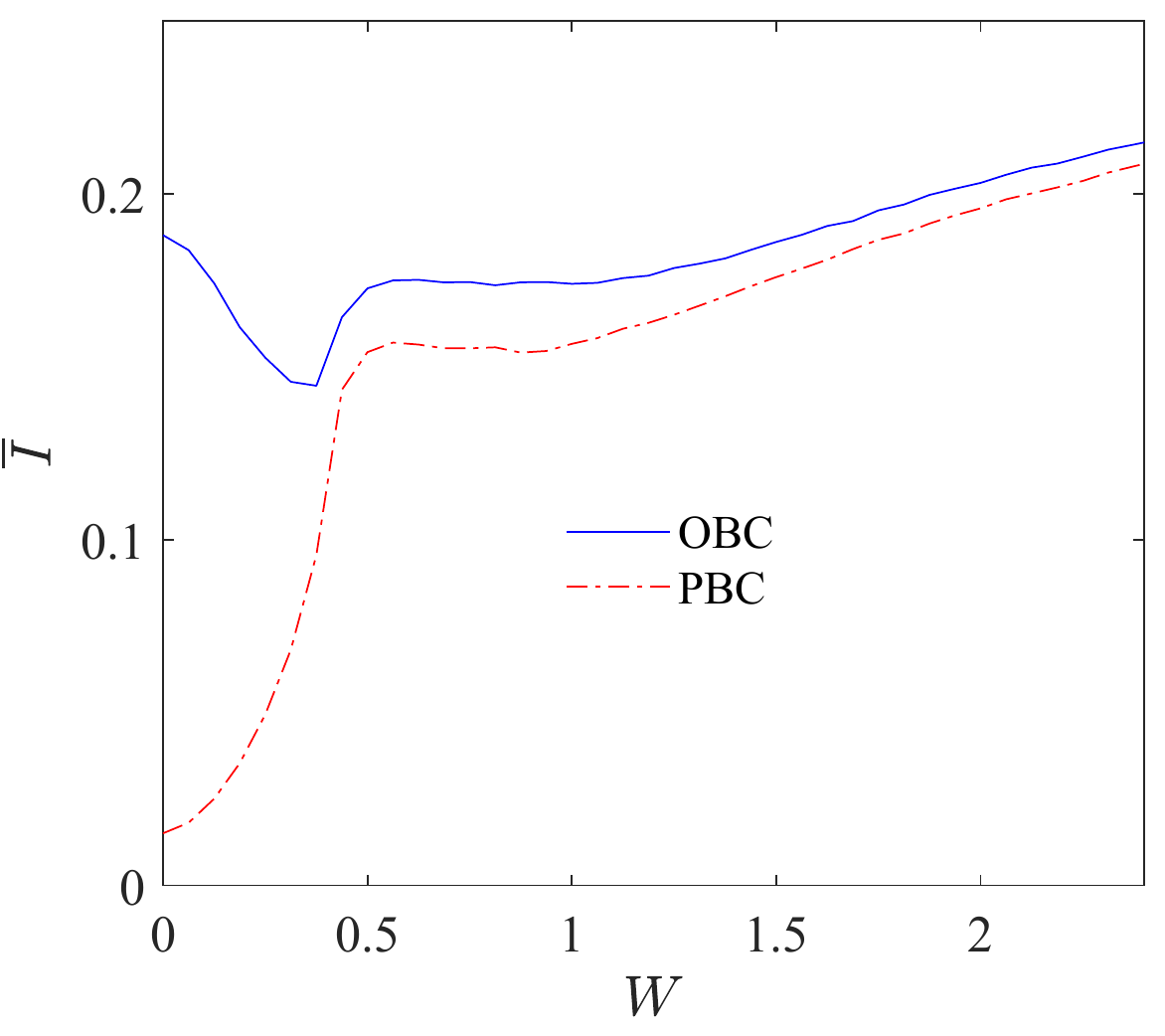}
\caption{Disorder-averaged IPR $\overline{I}$ as a function of the disorder strength $W$. We take $ \theta_{1}=4.3$, $\theta_{2}=2.175$, $\gamma=0.2$, and $N_c=100$. The IPR is averaged over $1000$ disorder configurations.}
\label{fig:Sipr}
\end{figure}

To provide a crosscheck on our experimental observation, we numerically calculate the inverse participation ratio (IPR) to characterize the localization properties of the system. Denoting the $m$th eigenstate of the Floquet operator $U$ as $|\psi^{(m)}\rangle=\sum_{x,i} c^{(m)}_{x,i} |x,i\rangle$ ($i=H,V$), we write the corresponding IPR as~\cite{ZDW1}
\begin{align}
 I=\frac{1}{2N_c}\displaystyle{\sum_m\sum_{i=H,V}\sum_x}|c^{(m)}_{x,i}|^4,
\end{align}
where $N_c$ is the total number of unit cells (external spatial modes). In Fig.~\ref{fig:Sipr}, we show the disorder-averaged IPR $\overline{I}$, where an additional average over different disorder configurations is performed. The calculated IPR shows a non-monotonic behavior with increasing disorder strength $W$, confirming the competition between the non-Hermitian skin effects and Anderson localization.

\section{Non-Bloch topology of the quantum walk without disorder}

Under the periodic boundary condition, the Floquet operator $U$ is topological, which preserves the chiral symmetry $\Gamma U \Gamma=U^{-1}$, with $\Gamma=\one_{w} \otimes \sigma_x$. Its winding number is given by
\begin{align}
\nu=\frac{1}{2\pi}\int dk \frac{-d_z\partial_{k} d_y+d_y\partial_{k} d_z}{d_y^2+d_z^2},
\label{eq:nu}
\end{align}
where the Bloch vector $\bm{d}=\text{Tr} (H\bm{\sigma})/|\text{Tr} (H\bm{\sigma})|$ [$k\in\left[0,2\pi\right)$]. Here the effective Hamiltonian $H$ is defined through $U=e^{-iH}$, and $\bm{\sigma}=(\sigma_x,\sigma_y,\sigma_z)$ ($\sigma_{x,y,z}$ are the Pauli matrices).

\begin{figure}
\centering
\includegraphics[width=0.4\textwidth]{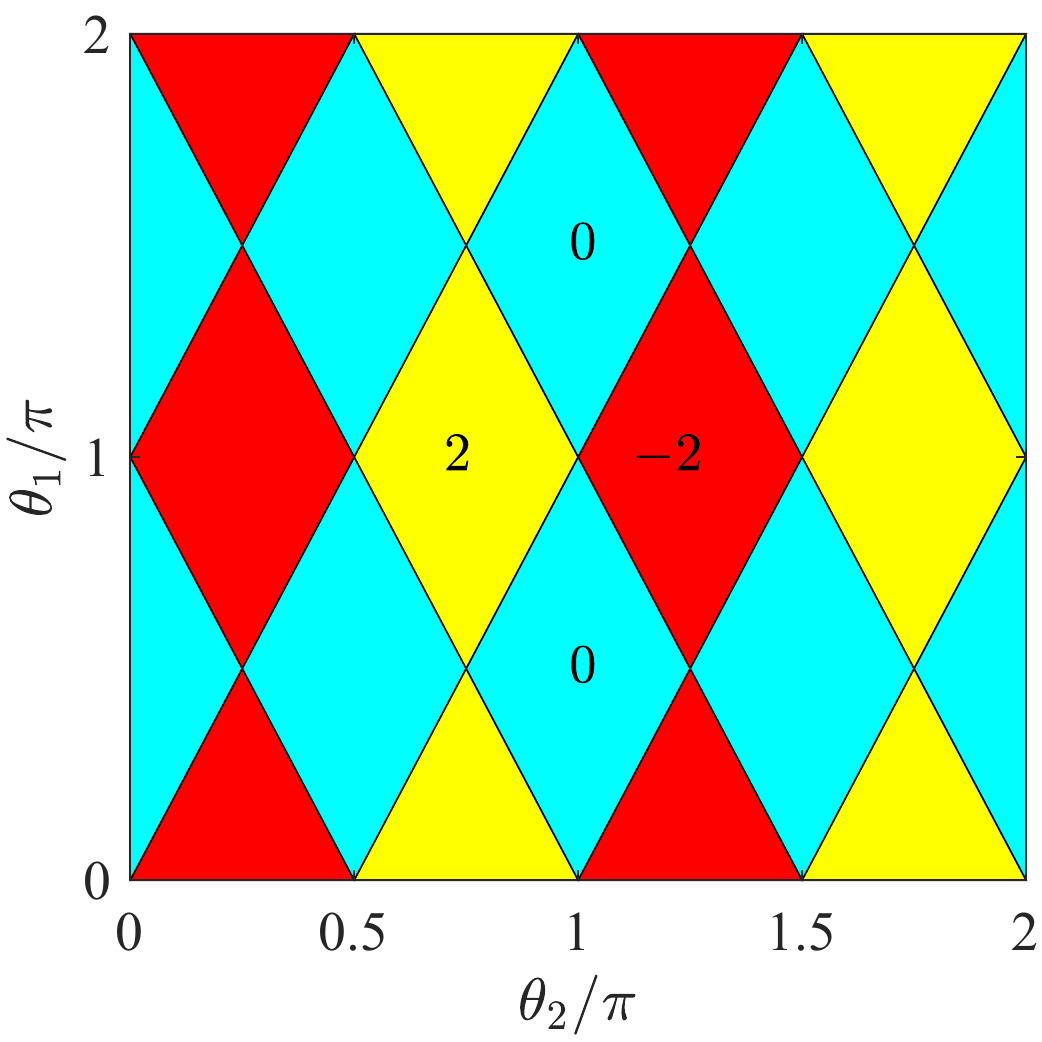}
\caption{Topological phase diagram in the parameter space $\theta_1$-$\theta_2$ characterized by the non-Bloch winding number $0$, $\pm 2$, under the open boundary condition.}\label{fig:Sphase}
\end{figure}

Under the open boundary condition, the topology is captured by the non-Bloch winding number. For that purpose, we replace $e^{ik}$ with $\beta=|\beta|e^{ik_\beta}$ [$k_\beta\in \left[0,2\pi\right)$] in Eq.~(\ref{eq:nu}), and perform the integration over $k_\beta$ to get the non-Bloch winding number. In Fig.~\ref{fig:Sphase}, we show the topological phase diagram in the absence of disorder. Incidentally, for our particular choice of $U$, the non-Bloch and Bloch phase boundaries are the same (thus independent of $\gamma$), despite the presence of the non-Hermitian skin effects and the deviation of the GBZ from the BZ.

\begin{figure}
\centering
\includegraphics[width=0.8\textwidth]{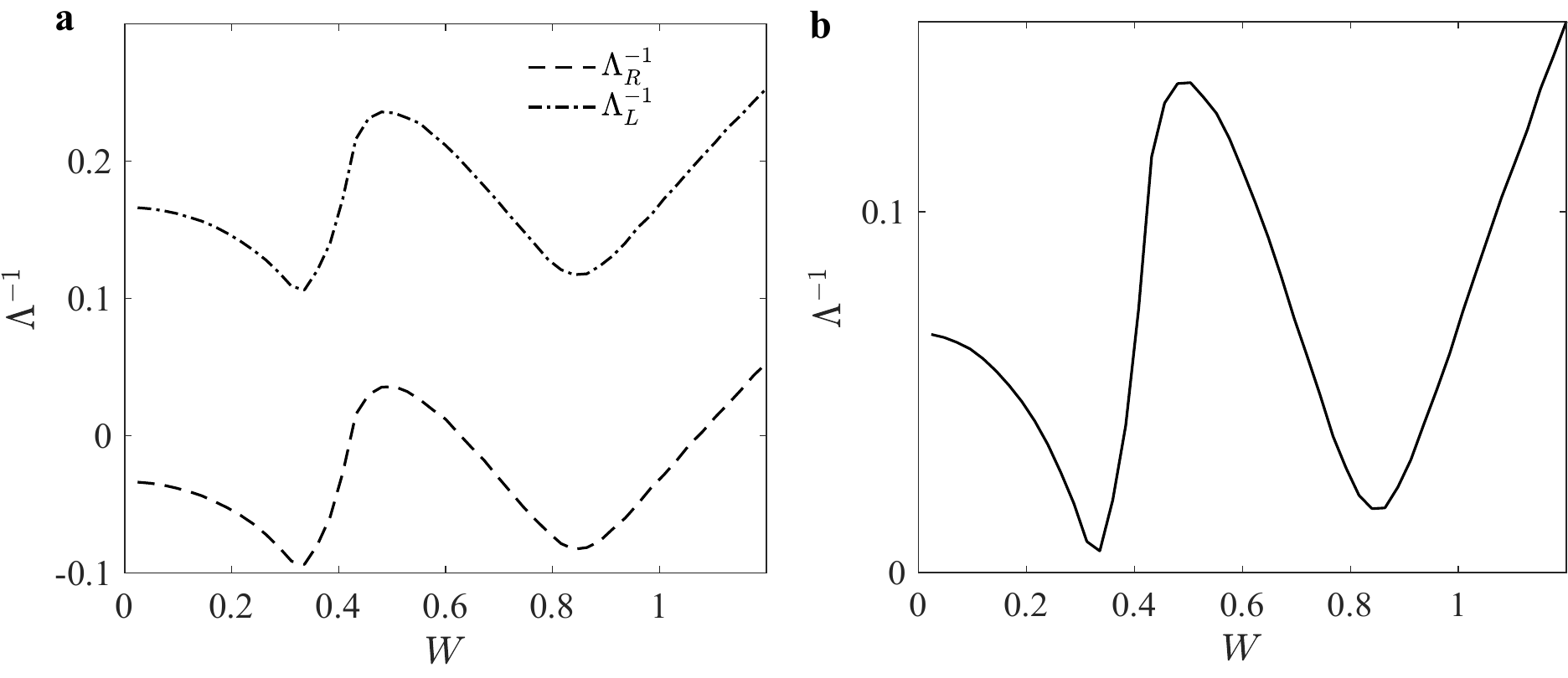}
\caption{Inverse localization lengths $\Lambda_{R,L}^{-1}$ in {\bf a}  and $\Lambda^{-1}$ in {\bf b} with increasing $W$.
We take the parameters $ \theta_{1}=4.3$, $\theta_{2}=2.175$, $\gamma=0.1$, and calculate the localization length of the $\pi$-quasienergy modes ($\lambda=-1$). For our calculations, the transfer matrix is iterated over $m=10^6$ times.}
\label{fig:Slocal}
\end{figure}

\begin{figure}
\includegraphics[width=0.5\textwidth]{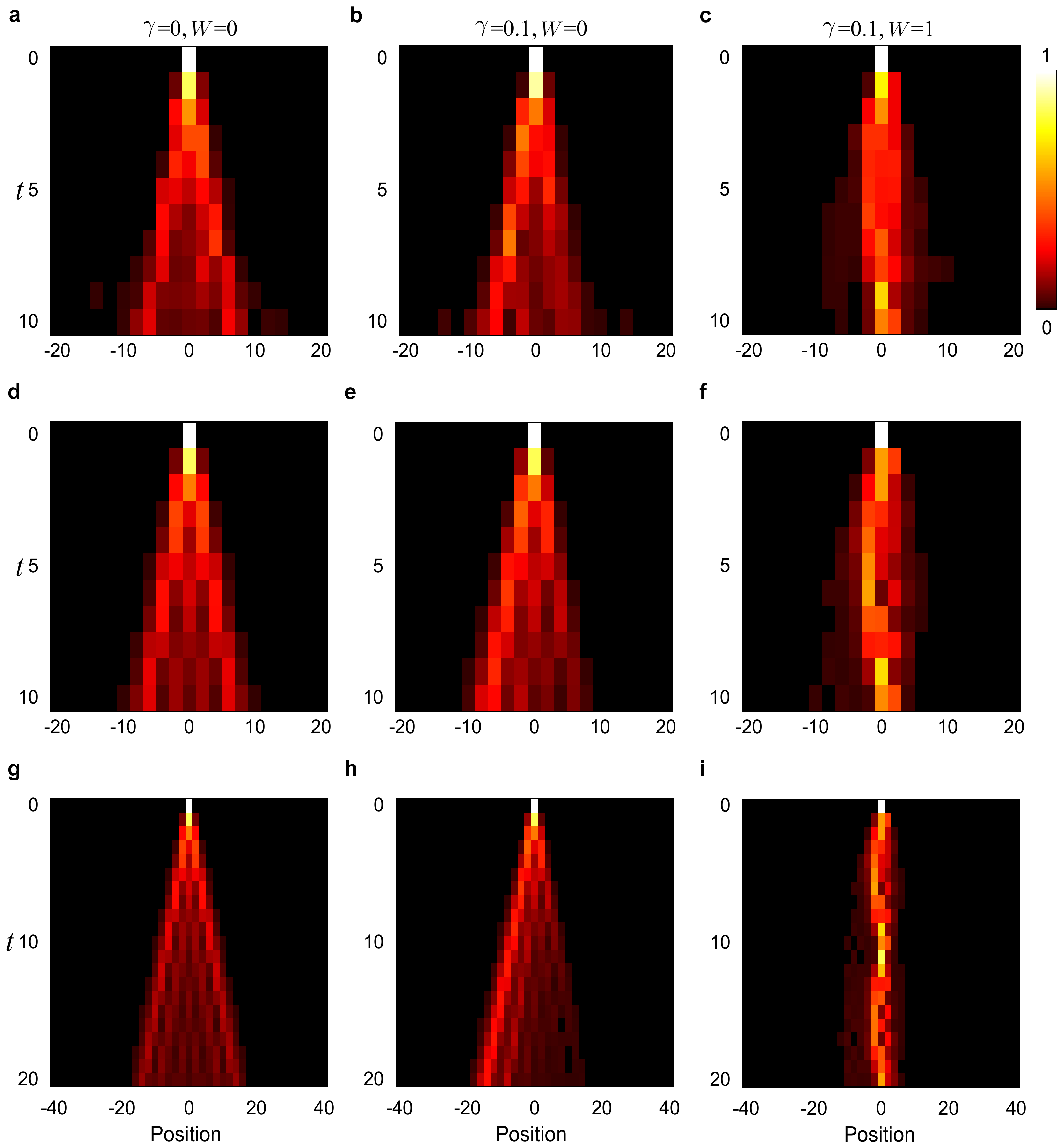}
\caption{{\bf a}, {\bf b}, {\bf c}, Measured polarization-averaged probability distribution under different loss and disorder parameters. The probability distribution is averaged over two independent $10$-step quantum walks with the initial states $|0\rangle\otimes|H\rangle$ and $|0\rangle\otimes|V\rangle$, respectively. The coin parameters are $\theta_1=4.3$ and $\theta_2=2.175$. {\bf d}, {\bf e}, {\bf f}, Numerically simulated polarization-averaged probability distribution following a $10$-step quantum walk, with parameters corresponding to those in {\bf a}, {\bf b} and {\bf c}. {\bf g}, {\bf h}, {\bf i}, Simulated polarization-averaged probability distribution following a $20$-step quantum walk, with the same parameters as those of {\bf d}, {\bf e} and {\bf f}. In {\bf c}, {\bf f}, and {\bf i}, only one disorder configuration is taken.
}
\label{fig:Sexp}
\end{figure}

\section{Biorthogonal local marker and chiral displacement}

Following Refs.~\cite{sciencecd,ZCW}, the biorthogonal local marker is defined as
\begin{equation}
\nu(m)=\frac{1}{4}\sum_s \langle m,s| Q \Gamma \left[X,Q\right]|m,s\rangle+h.c.,
\end{equation}
where $|m,s\rangle$ is the sublattice state $s$ of the $m$th unit cell, and $X$ is the unit-cell position operator. The biorthogonal projection operator $Q=P_{+}-P_{-}$, with $P_{\pm}= \sum_{n} | \phi^{(n)}_{\pm} \rangle \langle \chi^{(n)}_{\pm}|$. Where $|\phi^{(n)}_{\pm} \rangle$ is the $n$th right eigenstate of $U$, satisfying $U|\phi^{(n)}_{\pm} \rangle=\lambda^{(n)}_{\pm} |\phi^{(n)}_{\pm} \rangle$; and $ \langle\chi^{(n)}_{\pm} |$ is the $n$th left eigenstate, with $U^\dagger \ket{ \chi^{(n)}_{\pm}}=\lambda_{n,\pm}^*\ket{ \chi^{(n)}_{\pm}} $.
Here $\lambda_{n,+}$ ($\lambda_{n,-}$) lies in the lower (upper) half of the complex plane. Similar to the analysis in Refs.~\cite{sciencecd,ZCW}, the biorthogonal local marker serves as the topological invariant in a disordered system, and is reflected in the disorder- and time-averaged chiral displacement defined in Eq.~(\ref{eq:C}) of the main text.

\section{Biorthogonal localization length}

We now show how the biorthogonal localization length is calculated. As demonstrated in Fig.~\ref{fig:fig4}{\bf a} of the main text, the divergence of the biorthogonal localization length overlaps the topological phase boundary, suggesting its biorthogonal criticality.

Starting from the Floquet operator $U$ in Eq.~(\ref{eq:U}) of the main text, we write
$\theta_{2,x}= \theta_{2} $ and $\theta_{1,x}= \theta_{1}+ \delta_{x} $, where $ \delta_{x} \in \left[-W, W\right]$ is the uniformly distributed random disorder.

Transforming $U$ into the real space (spatial lattice modes), we have
\begin{align}
U=&\sum_{x}A_x|x-1,0\rangle \langle x+1,0| +B_x|x-1,0\rangle \langle x+1,1| +C_x|x,0\rangle \langle x,0| +D_x|x,0\rangle \langle x,1| \nonumber\\
&+E_x|x+1,0\rangle \langle x-1,0| +F_x|x+1,0\rangle \langle x-1,1| +G_x|x-1,1\rangle \langle x+1,0| +H_x|x-1,1\rangle \langle x+1,1| \nonumber\\
&+I_x|x,1\rangle \langle x,0| +J_x|x,1\rangle \langle x,1| +K_x|x+1,1\rangle \langle x-1,0|+L_x|x+1,1\rangle \langle x-1,1|
\end{align}
with the coefficients
\begin{align}
&A_x=e^{2\gamma}\cos^2\theta_2 \cos\theta_{1,x},  \\
&B_x=-e^{2\gamma}\cos\theta_2 \sin\theta_2\cos\theta_{1,x},\\
&C_x=-\cos\theta_2\sin\theta_2(\cos\theta_{1,x-1}+\cos\theta_{1,x+1}), \\ &D_x=\sin^2\theta_2\sin\theta_{1,x-1} -\cos^2\theta_2\sin\theta_{1,x+1},\\
&E_x=-e^{-2\gamma}\sin^2\theta_2\cos\theta_{1,x}, \\ &F_x=-e^{-2\gamma}\sin\theta_2\cos\theta_2\cos\theta_{1,x},\\
&G_x=e^{2\gamma}\sin\theta_2\cos\theta_2\cos\theta_{1,x}, \\  &H_x=-e^{2\gamma}\sin^2\theta_2\cos\theta_{1,x},\\
&I_x=\cos^2\theta_2\sin\theta_{1,x-1}-\sin^2\theta_2\sin\theta_{1,x+1}, \\ &J_x=-\cos\theta_2\sin\theta_2(\sin\theta_{1,x-1}+\sin\theta_{1,x+1}),\\
&K_x=e^{-2\gamma}\cos\theta_2\sin\theta_2\cos\theta_{1,x},   \\ &L_x=e^{-2\gamma}\cos^2\theta_2\cos\theta_{1,x}.
\end{align}

Writing the right eigenstate of $U$ as $|\psi_R\rangle=\sum_x c_{x,0}|x,0\rangle+c_{x,1}|x,1\rangle$, we have from the right eigen equation $U|\psi_R\rangle=\lambda|\psi_R\rangle$ ($\lambda$ is the eigenvalue)
\begin{align}
&A_{x+1}c_{x+2,0} +B_{x+1}c_{x+2,1}+(C_{x}-\lambda)c_{x,0}+D_{x}c_{x,1}+E_{x-1}c_{x-2,0}+F_{x-1}c_{x-2,1}=0,\\
&G_{x+1}c_{x+2,0} +H_{x+1}c_{x+2,1}+I_{x}c_{x,0}+(J_{x}-\lambda)c_{x,1}+K_{x-1}c_{x-2,0}+L_{x-1}c_{x-2,1}=0.
\end{align}

Noticing that $\frac{A_x}{G_x}=\frac{B_x}{H_x}=\cot\theta_2$, and $\frac{E_x}{K_x}=\frac{F_x}{L_x}=-\tan\theta_2$,  we have
\begin{align}
\mathcal{M}
\left(
  \begin{array}{ccc}
  c_{x+2,0}\\
  c_{x+2,1}
  \end{array}
\right)   =
 \mathcal{N}
\left(
  \begin{array}{ccc}
  c_{x,0}\\
  c_{x,1}
  \end{array}
\right),
\end{align}
where
\begin{align}
\mathcal{M}=&
\left(
  \begin{array}{ccc}
  C_{x+2}-\lambda-I_{x+2}\cot\theta_2& D_{x+2}-(J_{x+2}-\lambda)\cot\theta_2\\
  G_{x+1}&H_{x+1}
  \end{array}
\right) ,  \\
 \mathcal{N}=&-
\left(
  \begin{array}{ccc}
  E_{x+1}-K_{x+1}\cot\theta_2& F_{x+1}-L_{x+1}\cot\theta_2\\
  I_{x}\sin^2\theta_2+(C_{x}-\lambda)\cos\theta_2\sin\theta_2&(J_{x}-\lambda)\sin^2\theta_2+D_{x}\cos\theta_2\sin\theta_2
  \end{array}
\right).
\end{align}

We subsequently derive in an iterative fashion
\begin{align}
\left(
  \begin{array}{ccc}
  c_{2m,0}\\
  c_{2m,1}
  \end{array}
\right)   =
 \mathcal{T}_R^m
\left(
  \begin{array}{ccc}
  c_{0,0}\\
  c_{0,1}
  \end{array}
\right),
\end{align}
where $\mathcal{T}_R=\mathcal{M}^{-1}\mathcal{N}$ is identified as the right transfer matrix.

Similarly, we start from the left eigen equation $U^\dag |\psi_L\rangle=\lambda|\psi_L\rangle$ ($|\psi_L\rangle$ being the left eigenstate of $U$), and derive the left transfer matrix $\mathcal{T}_L$.

We may then define the right localization length $\Lambda_{R}$ through
$\Lambda^{-1}_{R}=\lim_{m\rightarrow \infty} \ln|T_{R,m}|/2n$; and the left localization length $\Lambda_L$ through
$\Lambda^{-1}_{L}=\lim_{m\rightarrow \infty} \ln|T_{L,m}|/2n$, where $T_{R,m}$
and $T_{L,m}$ are respectively the largest eigenvalues of $\mathcal{T}^m_R$ and $\mathcal{T}^m_L$. Note that, since $U^\dagger(\gamma)=\Gamma U(-\gamma)\Gamma$, and the chiral symmetry operator $\Gamma=\sum_x\ket{x}\bra{x} \otimes \sigma_x$ is local, $\Lambda_L$ of $U(\gamma)$ is equal to $\Lambda_R$ of $U(-\gamma)$. The biorthogonal location length $\Lambda$ is then defined as $\Lambda^{-1}=(\Lambda_R^{-1}+\Lambda_L^{-1})/2$.

In Fig.~\ref{fig:Slocal}, we plot the numerically calculated inverse localization lengths $\Lambda_{R,L}^{-1}$ and $\Lambda^{-1}$. Only the biorthogonal localization length diverges at the topological phase boundary $W=0.32$ and $W=0.84$. Note that when increasing the size of the system ($n$), $\Lambda^{-1}$ approaches zero at $W=0.32$ and $W=0.84$ in Fig.~\ref{fig:Slocal}{\bf b}.


\section{Additional experimental data}

In Fig.~\ref{fig:Sexp}, we show the polarization-averaged probability distribution in the bulk, both from experimental measurements and numerical simulations. Comparing Figs.~\ref{fig:Sexp}{\bf a} and {\bf b}, we see that under the non-Hermitian skin effect, the probability distribution tends to flow to the left (or to the right under other parameters), due to the presence of a bulk current. For sufficiently large $W$, the probability becomes localized for all times. These observations are consistent with those in Figs.~\ref{fig:fig2} and \ref{fig:fig3} of the main text.


\begin{thebibliography}{99}
\bibitem{kane}Hasan, M. Z. \& Kane, C. L. Colloquium: Topological insulators. {\it Rev. Mod. Phys.} {\bf 82}, 3045 (2010).
\bibitem{Qi}Qi, X.-L. \& Zhang, S.-C. Topological insulators and superconductors. {\it Rev. Mod. Phys.} {\bf 83}, 1057 (2011).
\bibitem{shen}Li, J., Chu, R.-L., Jain, J. K. \& Shen, S.-Q. Topological Anderson insulator. \textit{Phys. Rev. Lett.} {\bf 102}, 136806 (2009).
\bibitem{been}Groth, C., Wimmer, M., Akhmerov, A., Tworzydło, J. \& Beenakker, C. Theory of the topological Anderson insulator. \textit{Phys. Rev. Lett.} {\bf 103}, 196805 (2009).
\bibitem{Franz}Guo, H.-M., Rosenberg, G., Refael, G. \& Franz, M. Topological Anderson insulator in three dimensions. \textit{Phys. Rev. Lett.} {\bf 105}, 216601 (2010).
\bibitem{Hughes}Mondragon-Shem, I., Hughes, T. L., Song, J. \& Prodan, E. Topological criticality in the chiral-symmetric AIII class at strong disorder. \textit{Phys. Rev. Lett.} {\bf 113}, 046802 (2014).
\bibitem{Lee}Lee, T. E. Anomalous edge state in a non-Hermitian lattice, \textit{Phy. Rev. Lett.} {\bf 16}, 133903 (2016).
\bibitem{WZ1}Yao, S. \& Wang, Z. Edge states and topological invariants of non-Hermitian systems. \textit{Phys. Rev. Lett.} {\bf 121}, 086803 (2018).
\bibitem{WZ2}Yao, S., Song, F. \& Wang, Z. Non-Hermitian Chern bands. \textit{Phys. Rev. Lett.} {\bf 121}, 136802 (2018).
\bibitem{murakami}Yokomizo, K. \& Murakami, S. Non-Bloch band theory of non-Hermitian systems. \textit{Phys. Rev. Lett.} {\bf 123}, 066404 (2019).
\bibitem{ThomalePRB}Lee, C. H. \& Thomale, R. Anatomy of skin modes and topology in non-Hermitian systems. {\it Phys. Rev. B} {\bf 99}, 201103 (2019).
\bibitem{XZ}Zhang, X. \& Gong, J. Non-Hermitian Floquet topological phases: Exceptional points, coalescent edge modes, and the skin effect. \textit{Phys. Rev. B} {\bf 101}, 045415 (2020).
\bibitem{Budich}Kunst, F. K., Edvardsson, E., Budich, J. C. \& Bergholtz, E. J. Biorthogonal bulk-boundary correspondence in non-Hermitian systems. \textit{Phys. Rev. Lett.} {\bf 121}, 026808 (2018).
\bibitem{Slager}Borgnia, D. S., Kruchkov, A. J. \& Slager, R.-J. Non-Hermitian boundary modes and topology. \textit{Phys. Rev. Lett.} {\bf 124}, 056802 (2020).
\bibitem{mcdonald}McDonald, A., Pereg-Barnea, T. \& Clerk, A. Phase-dependent chiral transport and effective non-Hermitian dynamics in a Bosonic Kitaev-Majorana chain. \textit{Phys. Rev. X} {\bf 8}, 041031 (2018).
\bibitem{alvarez}Alvarez, V. M., Vargas, J. B. \& Torres, L. F. Non-Hermitian robust edge states in one dimension: Anomalous localization and eigenspace condensation at exceptional points. \textit{Phys. Rev. B} {\bf 97}, 121401 (2018).
\bibitem{fangchenskin}Zhang, K., Yang, Z. \& Fang, C. Correspondence between winding numbers and skin modes in non-Hermitian systems. \textit{Phys. Rev. Lett.} {\bf 125}, 126402 (2020).
\bibitem{kawabataskin}Okuma, N., Kawabata, K., Shiozaki, K. \& Sato, M. Topological origin of non-Hermitian skin effects. \textit{Phys. Rev. Lett.} {\bf 124}, 086801 (2020).
\bibitem{yzsgbz}Yang, Z., Zhang, K., Fang, C. \& Hu, J. Non-Hermitian bulk-boundary correspondence and auxiliary generalized Brillouin zone theory. \textit{Phys. Rev. Lett.} {\bf 125}, 226402 (2020).
\bibitem{stefano}Longhi, S. Probing non-Hermitian skin effect and non-Bloch phase transitions. \textit{Phys. Rev. Research} {\bf 1}, 023013 (2019).
\bibitem{tianshu}Deng, T.-S. \& Yi, W. Non-Bloch topological invariants in a non-Hermitian domain wall system. \textit{Phys. Rev. B} {\bf 100}, 035102 (2019).
\bibitem{lli}Li, L., Lee, C. H., Mu, S. \& Gong, J. Critical non-Hermitian skin effect, \textit{Nat. Commun.} {\bf 11}, 1-8 (2020).
\bibitem{Szameit2}St\"{u}tzer, S. \textit{et al.} Photonic topological Anderson insulators, \textit{Nature} {\bf 560}, 461-465 (2018).
\bibitem{sciencecd}Meier, E. J. \textit{et al.} Observation of the topological Anderson insulator in disordered atomic wires. \textit{Science} {\bf 362}, 929 (2018).
\bibitem{Szameit}Weidemann, S., Kremer, M., Longhi, S. \& Szameit, A. Coexistence of dynamical delocalization and spectral localization through stochastic dissipation. \textit{Nat. Photon.} 1-6 (2021).
\bibitem{Szameit1}Maczewsky, L. J. \textit{et al.} Nonlinearity-induced photonic topological insulator. \textit{Science} {\bf 370}, 701-704 (2020).
\bibitem{teskin}Helbig, T. \textit{et al.} Generalized bulk–boundary correspondence in non-Hermitian topolectrical circuits. \textit{Nat. Phys.} {\bf 16}, 747 (2020).
\bibitem{photonskin}Xiao, L. \textit{et al.} Observation of non-Hermitian bulk-boundary correspondence in quantum dynamics. \textit{Nat. Phys.} {\bf 16}, 761 (2020).
\bibitem{metaskin}Ghatak, A., Brandenbourger, M., van Wezel, J. \& Coulais, C. Observation of non-Hermitian topology and its bulk-edge correspondence in an active mechanical metamaterial. \textit{Proc. Natl. Ac. Sc.} {\bf 117}, 29561 (2020).
\bibitem{teskin2d}Hofmann, T. \textit{et al.} Reciprocal skin effect and its realization in a topolectrical circuit. \textit{Phys. Rev. Research} {\bf 2}, 023265 (2020).
\bibitem{scienceskin}Weidemann, S. \textit{et al.} Topological funneling of light. \textit{Science} {\bf 368}, 311 (2020).
\bibitem{dzou}Zou, D. \textit{et al.} Observation of hybrid higher-order skin-topological effect in non-Hermitian topolectrical circuits. arXiv: 2104.11260.
\bibitem{EP}Xiao, L. \textit{et al.} Observation of non-Bloch parity-time symmetry and exceptional points. \textit{Phys. Rev. Lett.} {\bf 126}, 230402 (2021).
\bibitem{ZDW1}Zhang, D.-W., Tang, L.-Z., Lang, L.-J., Yan, H. \& Zhu, S.-L. Non-Hermitian topological Anderson insulators. \textit{Sci. China-Phys. Mech. Astron.} {\bf 63}, 267062 (2020).
\bibitem{ZCW}Luo, X.-W. \& Zhang, C. Non-Hermitian disorder-induced topological insulators. arXiv: 1912.10652.
\bibitem{ZDW2}Tang, L.-Z., Zhang, L.-F., Zhang, G.-Q. \& Zhang, D.-W. Topological Anderson insulators in two-dimensional non-Hermitian disordered systems. \textit{Phys. Rev. A} {\bf 101}, 063612 (2020).
\bibitem{hug}Claes, J. \& Hughes, T. L. Skin effect and winding number in disordered non-Hermitian systems. \textit{Phys. Rev. B} {\bf 103}, 140201 (2021).
\bibitem{Silb1}Schreiber, A. \textit{et al.} Photons walking the line: a quantum walk with adjustable coin operations. \textit{Phys. Rev. Lett.} {\bf 104}, 050502 (2010).
\bibitem{Silb2}Schreiber, A. \textit{et al.} Decoherence and disorder in quantum walks: from ballistic spread to localization. \textit{Phys. Rev. Lett.} {\bf 106}, 180403 (2011).
\bibitem{lu}Chen, C. \textit{et al.} Observation of topologically protected edge states in a photonic two-dimensional quantum walk. \textit{Phys. Rev. Lett.} {\bf 121}, 100502 (2018).

\end{thebibliography}
\end{document}